\documentclass[11pt,twoside]{article}

\usepackage{asp2004}
\usepackage{epsf}
\usepackage{psfig}
\usepackage{lscape}
\usepackage{url}
\begin{document}
\title{Spectral analyses of 16 DAO white dwarfs from the Sloan Digital
  Sky Survey}   
\author{S.~D.~H\"ugelmeyer$^1$, S.~Dreizler$^1$, T.~Rauch$^2$, and
  J.~Krzesi\'nski$^3$}
\affil{$^1$Institut f\"ur Astrophysik, Universit\"at$\,$G\"ottingen,
  Friedrich-Hund-Platz~1, D-37077 G\"ottingen, Germany \\
  $^2$Institut f\"ur Astronomie und Astrophysik,
  Universit\"at$\,$T\"ubingen, Sand~1, D-72076 T\"ubingen, Germany\\
  $^3$New Mexico State University, Apache Point Observatory, 2001
  Apache Point Road, P.O. Box 59, Sunspot, NM 88349, USA}

\begin{abstract} 
We present a spectral analysis of 16 DAO white dwarfs from the Sloan
Digital Sky Survey Data Release 4. With our NLTE H+He model grid, we
derived photospheric parameters for these objects. We compare our new
results to literature values and divide the DAOs into two distinct
groups: post-AGB and EHB progenitors.
\end{abstract}


\section{Introduction} 

Even though most white dwarfs (WDs) have a simple atmospheric composition,
either H- or He-dominated, some objects show mixed
compositions, for example DAO WDs. These stars play an
important role in the understanding of the evolution and nature of
white dwarfs. Typical atmospheric parameters for DAOs are $T_{\rm
eff}=50-100$~kK, $\log[{\rm He/H}] \sim -2$ (by number), and a 
low surface gravity of $\log{g} \la 7.5$ (cgs)
\citep{1994ApJ...432..305B,1999A&A...350..101N,2004MNRAS.355.1031G}.

The evolution towards DAO white dwarfs is still not satisfyingly
understood. Because of the low surface gravity and low mass ($M <
0.5~M_\odot$), derived for a large part of the analysed DAOs, a
post-asymptotic giant branch (post-AGB) evolution of these
objects seems to be rather unlikely since the mass is not high enough
to ignite helium-shell burning on the horizontal branch (HB). Instead,
an extended horizontal branch (EHB) history for these low-mass WDs is
considered.

For those DAOs which are massive enough to start helium-shell burning
on the HB, i.e.\ $M>0.53~M_\odot$, a post-AGB
evolution is assumed. The star either evolves from a hydrogen-rich
post-AGB star or a (hybrid) PG~1159 star. In the first case, one would
expect a (sub-)solar helium abundance in the atmosphere of the
DAO while the evolution from a hydrogen-deficient post-AGB object is
realised by the decrease of mass loss and the resulting up-floating of
hydrogen.

\begin{table}[!ht]
  \caption{Atmospheric parameters of DAO white dwarfs. Masses are derived from
  evolution tracks of \citet{1995LNP...443...41W} assuming a thick
  hydrogen layer. Uncertanties are given by 1-$\sigma$ statistical
  errors. \label{tab:dao}}
  \vspace{-.1in}
  \begin{center}
    {\footnotesize
      \begin{tabular}{l r @{\,$\pm$\,} l r @{\,$\pm$\,} l r
	  @{\,$\pm$\,} l r @{\,$\pm$\,} l}
	\hline\hline\noalign{\smallskip}
	Name & \multicolumn{2}{c}{$T_{\rm eff}$} &
	\multicolumn{2}{c}{$\log g$} & \multicolumn{2}{c}{$\log[{\rm
	He/H}]$} & \multicolumn{2}{c}{$M$} \\
	SDSS J&  \multicolumn{2}{c}{[kK]} & \multicolumn{2}{c}{(cgs)}
	& \multicolumn{2}{c}{(by numer)} &
	\multicolumn{2}{c}{[$M_\odot$]} \\
	\noalign{\smallskip}
	\hline
	145606.73+491116.5$^a$   &90.4&2.2 & 6.57&0.07 & $-$0.970&0.001 & 0.44&0.01\\
	121743.14+623118.3       &87.5&1.8 & 6.80&0.08 & $-$0.929&0.001 & 0.46&0.01\\
	160236.08+381950.6       &87.1&3.5 & 6.63&0.08 & $-$1.725&0.004 & 0.43&0.01\\
	131925.93+531715.0       &83.5&2.9 & 6.57&0.09 & $-$0.742&0.001 & 0.41&0.01\\
	034831.34+004616.3$^b$   &79.2&1.1 & 6.99&0.07 & $-$1.982&0.014 & 0.47&0.01\\
	120927.95$-$030206.3$^c$ &75.3&2.6 & 6.78&0.09 & $-$1.725&0.004 & 0.41&0.02\\
	153102.41+534900.6       &74.8&3.4 & 6.48&0.10 & $-$1.881&0.009 & 0.34&0.02\\
	125029.51+505317.4       &69.3&5.8 & 6.56&0.20 & $-$1.177&0.012 & 0.32&0.05\\
	082705.53+313008.3$^d$   &67.8&0.7 & 6.84&0.05 & $-$1.881&0.009 & 0.38&0.01\\
	163200.32$-$001928.3$^e$ &64.5&4.5 & 7.86&0.13 & $-$1.079&0.001 & 0.66&0.05\\
	101015.60+115711.3       &59.5&1.6 & 8.10&0.13 & $-$0.910&0.004 & 0.76&0.07\\ 
	161441.99+370548.2       &59.0&0.6 & 7.76&0.07 & $-$1.982&0.014 & 0.61&0.03\\
	135356.89$-$025630.4     &50.7&1.3 & 7.86&0.06 & $-$1.467&0.001 & 0.63&0.03\\
	081618.80+034234.2       &50.0&0.5 & 7.00&0.09 & $-$0.970&0.001 & 0.36&0.03\\
	235137.25+010844.1       &50.0&0.3 & 7.76&0.06 & $-$0.294&0.001 & 0.59&0.02\\
	170508.82+212019.3       &50.0&0.1 & 8.04&0.04 & $-$1.982&0.014 & 0.71&0.02\\
	\hline
	\noalign{\smallskip}
	\multicolumn{9}{l}{{\footnotesize $^a$: PG\,1454+494: sd
	  \citep{1988SAAOC..12....1K}; $^b$: KUV\,03459+0037: sdB }} \\
	\multicolumn{9}{l}{{\footnotesize \citep{1993AJ....105..660W};
	    $^c$: PG\,1206$-$028: sdO(B) \citep{1988SAAOC..12....1K};}} \\
	\multicolumn{9}{l}{{\footnotesize  $^d$: TON~320: DAO
	    \citep{1994ApJ...432..305B}; $^e$: WD\,1629$-$002: DAO}} \\
	\multicolumn{9}{l}{{\footnotesize \citep{2004A&A...417.1093K}}} \\
      \end{tabular}
    }
  \end{center}
\end{table}

\section{Models and fitting}

We used NLTE synthetic spectra calculated with TMAP
\citep{2003WSAM...W,2003WSAM...R} from an extended atmosphere grid of
H+He composed models with H/He abundance ratios of 10$^9$:1, 9:1, 8:2,
7:3, \dots, 1:9, 1:10$^9$ by mass. Furthermore, we interpolated models
with H/He abundance ratios of 9.5:0.5, 8.5:1.5, and 7.5:2.5 by
mass. The model grid ranges from $50 - 190$~kK and $\log{g}$ is in the
region from $5.0 - 9.0$. The emergent fluxes were calculated using
Stark broadening tables of \citet{1997A&AS..122..285L},
\citet{1974JQSRT..14.1025B}, and \citet{1989A&AS...78...51S} for H I,
He I, and He II line broadening, respectively. The model atmospheres
were calculated for a homogeneous composition which is appropriate for
DAOs as shown by \citet{1994ApJ...432..305B}.

\begin{figure}
  \plotone{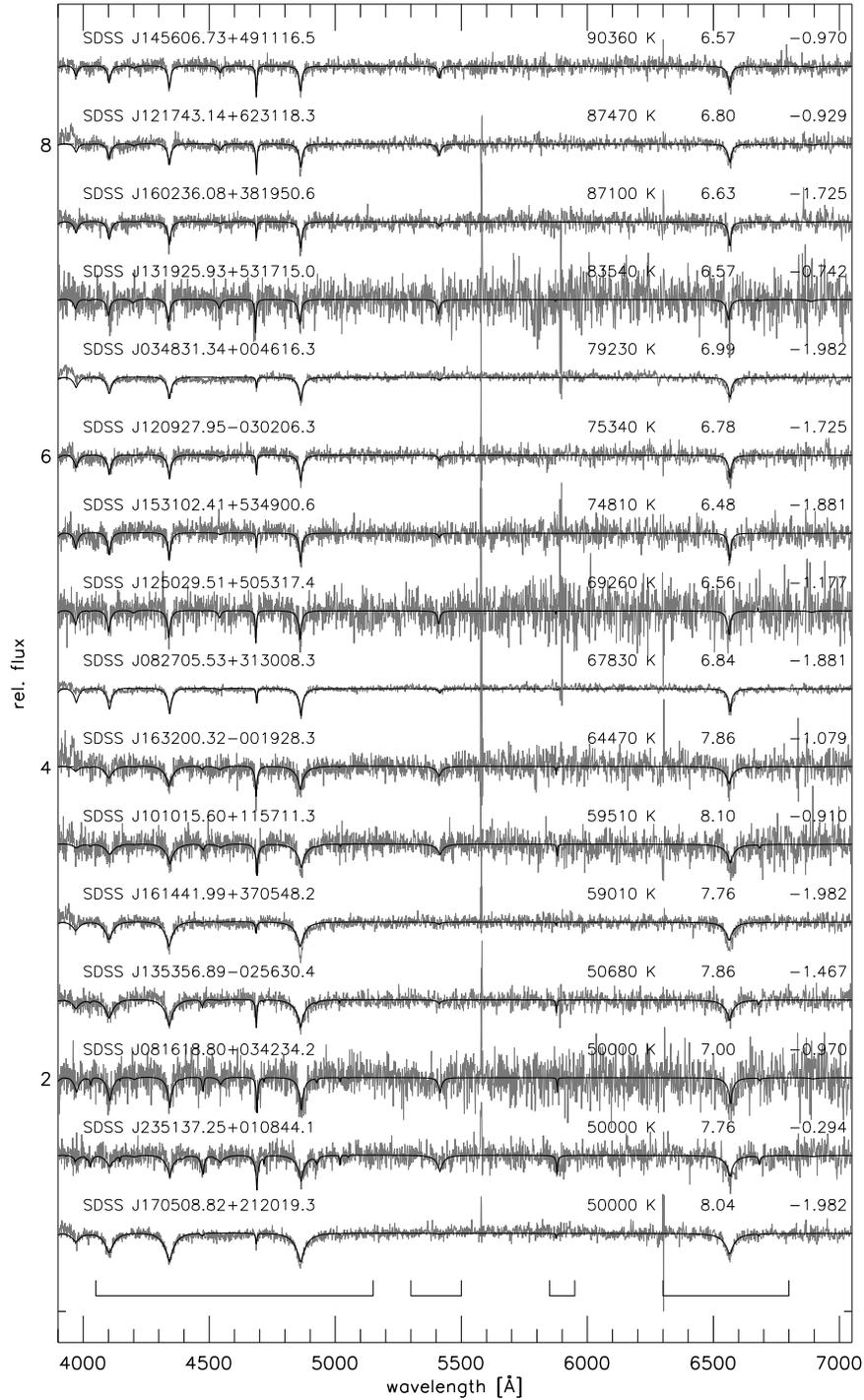}
  \caption{SDSS spectra (grey lines) and best-fit models (black lines)
    of DAO white dwarfs. The labels denote SDSS name, $T_{\rm eff}$,
    $\log{g}$, and $\log[{\rm He/H}]$. The marks at the bottom show
    regions where fitting has been performed. \label{fig:daos}}
\end{figure}

We determined best-fit models using our automated $\chi^2$-fitting
routines \citep{2006...diploma}. We fitted the spectrum in the regions
marked in Fig.~\ref{fig:daos} and performed an interpolation between
$\chi^2$-values to refine photospheric
parameters. \citet{1999A&A...350..101N} reports that the high Balmer
lines are the most solid temperature indicator. Our fitting yields
best-fit models that reproduce H$\gamma$ and H$\delta$ most
accurately. Three DAOs hit the lower $T_{\rm eff}$ limit of our model
grid. In these cases one-sided errors are given. The SDSS spectra
together with best-fit models are shown in Fig.~\ref{fig:daos}.

\section{Results and discussion}

\begin{figure}
  \plotone{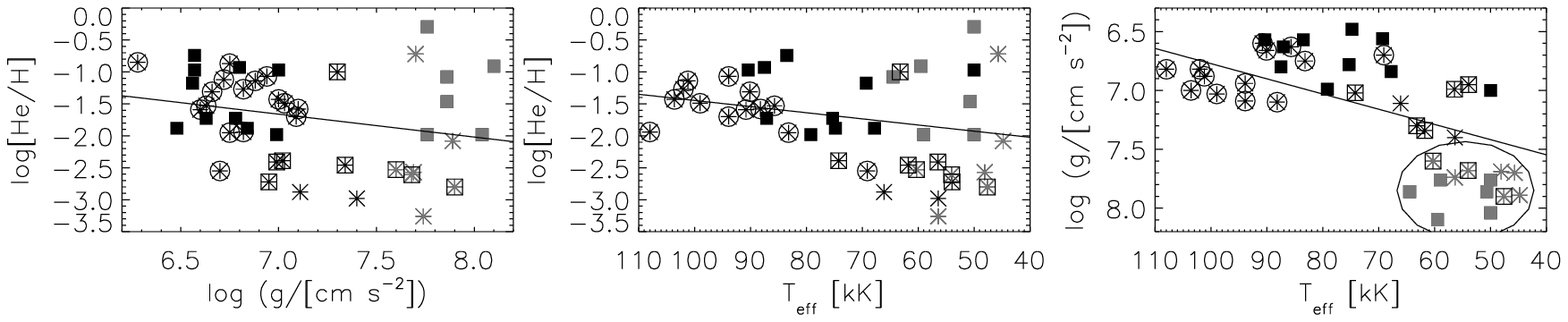}
    \caption{Photospheric parameters for known DAO white
    dwarfs. Literature values are taken from
    \citet[][asterisks]{1994ApJ...432..305B},  
    \citet[][circled asterisks]{1999A&A...350..101N}, and
    \citet[][squared asterisks]{2004MNRAS.355.1031G}
    and this work (filled squares). The line is a linear fit
    through all data points. \label{fig:dao_para}}
\end{figure}

The atmospheric parameters derived from our analysis are shown in
Table~\ref{tab:dao} and, compared to literature values from
\citet{1994ApJ...432..305B}, \citet{1999A&A...350..101N}, and
\citet{2004MNRAS.355.1031G}, in the plots of Fig.~\ref{fig:dao_para}. It
suggests a division of the DAOs into two distinct groups: all
stars represented by grey symbols have $\log{g} > 7.5$ and
a mass $> 0.5~M_\odot$, which is sufficient for the progenitor HB
star to start He-shell burning and ascend the AGB. Therefore,
these stars should have had a post-AGB history. The black symbols in
Fig.~\ref{fig:dao_para} are less compact objects among which we find the
H-rich CSPNs from \citet{1999A&A...350..101N} ($M >
0.5~M_\odot$ except for GD~561 or Sh 2-174 with $M = 0.43~M_\odot$)
and the less massive DAOs which we believe evolved directly from the
EHB or in a binary evolution directly from the RGB
\citep{1998A&A...339..123D}. This division in post-AGB and EHB
progenitors is further underlined by the fact that the high gravity
objects also have a tendency to be more He-rich than the other
DAOs which implies that the former ones originate from He-rich
post-AGB stars such as (hybrid) PG~1159 stars.

We have excluded the possibility that the mixed hydrogen and helium
spectra originate from double-degenerate binary systems consisting of
e.g.\ a DA and a DO white dwarf. Radial velocity measurements are
necessary to check for binarity of the SDSS DAO white dwarfs of this
analysis.

\acknowledgements S.D.H. would like to thank the Royal Astronomical Society
and the Berliner-Ungewitter-Stiftung for generous financial support.

\end{document}